\documentstyle{llncs}

\begin{document}

\pagestyle{empty}


\title{AN ALGORITHM TO VERIFY LOCAL THRESHOLD TESTABILITY OF DETERMINISTIC
FINITE AUTOMATA}

\author{A.N. Trahtman}

\institute{Bar-Ilan University, Dep. of Math. and CS, 52900,Ramat Gan, Israel}

\maketitle
\centerline{Lecture Notes in Computer Scence 2214(2001), 164-173}

\newtheorem{defn}{D\'efinition}[section]
\newtheorem{quest}[defn]{Question}
\newtheorem{prop}[defn]{Proposition}
\newtheorem{thm}[defn]{Theorem}
\newtheorem{lem}[defn]{Lemma}
\newtheorem{cor}[defn]{Corollary}


 email:trakht@macs.biu.ac.il


\begin{abstract} A locally threshold testable language $L$ is a language with
        the property that for some nonnegative integers $k$ and $l$,
         whether or not a word $u$ is in the language $L$ depends on
        (1) the prefix and suffix of the word $u$ of length $k-1$ and
        (2) the set of intermediate
        substrings of length $k$ of the word $u$ where the sets of substrings
        occurring {\em at least} $j$ times are the same, for $j \le l$.
        For given $k$ and $l$ the language is called $l$-threshold
         $k$-testable.
        A finite deterministic automaton is called $l$-threshold $k$-testable
        if the automaton accepts a $l$-threshold $k$-testable language.

  In this paper,
   the necessary and sufficient conditions for an
automaton to be locally threshold testable are found.
   We introduce the first polynomial time
   algorithm to verify local threshold testability of the automaton based
   on this characterization.

  New version of polynomial time algorithm to verify the local testability
 will be presented too.

  \end{abstract}



 \section*{Introduction}
    The concept of local testability was introduced by McNaughton
 and Papert $\cite {MP}$ and by Brzozowski and Simon $\cite {BS}$.
Local testability can be considered as a special case of local
 $l$-threshold testability for $l=1$.  Locally testable
 languages, automata and semigroups have been
 investigated from different points of view (see  $\cite {Bi}$ -
 $\cite {K94}$, $\cite {Ma}$, $\cite {Pi}$, $\cite {Tr}$ - $\cite {ZC}$).
 In $\cite {Mi}$, local testability was discussed in terms of
 "diameter-limited perceptrons". Locally testable languages are a
 generalization of the definite and reverse-definite languages, which can be
found, for example, in $\cite {G}$ and $\cite {Sh}$.
    Some variations of the concept of local testability (strictly, strongly)
    obtained by changing or omitting prefixes and suffixes in the
    definition of the concept were studied in $\cite {Bi}$, $\cite {C}$,
     $\cite {MP}$, $\cite {Pi}$, $\cite {Tr}$.

 Locally testable automata have a wide spectrum of applications.
  Regular languages and picture languages can be described by a strictly
  locally testable languages $\cite {Bi}$, $\cite {H}$.
 Local automata (a kind of locally testable automata) are heavily
used to construct transducers and coding schemes adapted to constrained
channels $\cite {Be}$.
Literal morphisms may be modelled by help of 2-testable languages
 $\cite {ES}$.
  In  $\cite {He}$, locally testable languages are used in the study of DNA
 and informational macromolecules in biology.

 Kim, McNaughton and McCloskey ($\cite {K89}$, $\cite {K91}$) have found
   necessary and sufficient conditions of local testability
and a polynomial time algorithm for
local testability problem based on these conditions.  The
realization of the algorithm is described by Caron in $\cite {C}$.
A polynomial time algorithm for
local testability problem for semigroups was presented in $\cite {Ts}$.

    The locally threshold testable languages were introduced by Beauquier and
 Pin $\cite {BP}$. These languages generalize the concept of
 locally testable language and have been studied extensively in recent years
 (see $\cite {BP1}$, $\cite {Pi}$,  $\cite {To}$, $\cite {W}$, $\cite {Wi}$),
 $\cite {REG}$.

An important reason to study locally threshold testable languages
is the possibility of being used in pattern recognition $\cite {Fu}$,
$\cite {GT}$, $\cite {REG}$. Stochastic
 locally threshold testable languages, also known as $\it N-grams$ are used
 in pattern recognition, particulary in speech recognation, both in
acoustic-phonetics decoding as in language modeling $\cite {VCG}$.
An application of these languages to the inference problem considered in
$\cite {REG}$.
   The syntactic characterization of locally threshold
 testable languages one can find in $\cite {BP}$:

    Given the syntactic
  semigroup $S$ of the language $L$, we form a graph G(S) as follows. The
  vertices of G(S) are the idempotents of $S$, and the edges from $e$ to $f$
   are the elements of the form $esf$. A language $L$ is locally threshold
   testable if and only if $S$ is aperiodic and for any two nodes $e$, $f$
   and three edges $p$, $q$, $r$ such that $p$ and $q$ are edges from $e$ to
   $f$ and $r$ is an edge from $f$ to $e$ we have

 \centerline{prq=qrp}

  Since only five elements of the semigroup $S$ are considered, there
 exists a polynomial time algorithm of order $O(|S|^5)$ for local threshold
testability problem in the case of semigroups.  But the cardinality of the
   syntactic semigroup of a locally threshold testable automaton is not
    polynomial in the number of its nodes $\cite {K91}$.  This is why the study
    of the automaton and the state transition graph of the automaton is
    important from the practical point of view (see $\cite {K91}$, $\cite
    {K94}$) and we use here this approach.

   For the state transition graph $\Gamma$ of an automaton, we consider some
subgraphs of the cartesian product $\Gamma \times \Gamma$ and
  $\Gamma \times \Gamma \times \Gamma$.
    In this way,
 necessary and sufficient conditions for a deterministic finite
automaton to be locally threshold testable are found.
 We present here $O(n^5)$ time
   algorithm to verify local threshold testability of the automaton based
   on this characterization.

   Necessary and sufficient conditions of local testability from $\cite {K91}$
  are considered in this paper in terms of reachability in the graph
 $\Gamma \times \Gamma$.
  New version of $O(n^2)$ time algorithm to verify local testability
 based on this approach will be presented too.

 \section*{Notation and definitions}

 Let $\Sigma$ be an alphabet and let $\Sigma^+$ denote  the free semigroup
on $\Sigma$. If $w \in \Sigma^+$, let $|w|$ denote the length of $w$.
Let $k$ be a positive integer. Let $i_k(w)$ $[t_k(w)]$ denote the prefix
[suffix] of $w$ of length $k$ or $w$ if $|w| < k$. Let $F_{k,j}(w)$ denote
 the set of factors of $w$ of length $k$ with at least $j$ occurrences.
   A language $L$ [a semigroup $S$]
 is called {\bf l-threshold k-testable} if there is an alphabet $\Sigma$
 [and a surjective morphism $\phi : \Sigma^+ \to S$] such that  for all
  $u$, $v \in \Sigma^+$, if $i_{k-1}(u)=i_{k-1}(v)$, $t_{k-1}(u)=t_{k-1}(v)$
 and $F_{k,j}(u)=F_{k,j}(v)$ for all $j \le l$, then either both $u$ and $v$
 are in $L$ or neither is in $L$ [$u\phi = v\phi$].

   An automaton is {\bf $l$-threshold $k$-testable} if the automaton accepts a
 $l$-threshold $k$-testable language [the syntactic semigroup of the automaton is
 {\bf $l$-threshold $k$-testable}].

   A language $L$ [a semigroup $S$, an automaton $\bf A$] is {\bf locally
 threshold testable} if it is $l$-threshold $k$-testable for some $k$ and
 $l$.

\medskip
 A semigroup without non-trivial subgroups is called {\bf aperiodic} $\cite
 {BP}$

$|\Gamma|$ denotes the number of nodes of the graph $\Gamma$.

$\Gamma^i$ denotes the direct product of $i$ copies of the graph $\Gamma$.

A maximal strongly connected component of the graph will be denoted
 for brevity as $\bf SCC$ $\cite {K89}$,
a finite deterministic automaton will be denoted as $\bf DFA$ $\cite {K91}$.
 A node from an $SCC$ will be called for brevity as an $\bf SCC-node$.

If an edge ${\bf p} \to \bf q$ is labeled by $\sigma$ then
let us denote the node $\bf q$ as ${\bf p}\sigma$.

 We shall write $\bf p \succeq \bf q$ if the node $\bf q$  is reachable from
the node $\bf p$ or $\bf p=q$.

In the case $\bf p \succeq q$ and $\bf q \succeq p$ we write $\bf p \sim q$
(that is $\bf p$ and $\bf q$ belong to one $SCC$ or $\bf p=q$).

 \section{The necessary and sufficient conditions}
   Let us formulate the result of Beauquier and Pin $\cite {BP}$ in the
   following form:
   \begin{thm} $\label {0.0}$ $\cite {BP}$
    A language $L$ is locally threshold
   testable if and only if the syntactic semigroup $S$ of $L$ is aperiodic
and for any two idempotents $e$, $f$ and elements $a$, $u$, $b$ of $S$ we have

    \begin{equation}
                    eafuebf=ebfueaf                      \label {1}
   \end{equation}
  \end{thm}
   \medskip

   Let us recall the concept of implicit operation $\cite {R}$, $\cite {Al}$:
     The unary
    operation $x^{\omega}$ assigns to every element $x$ of a finite semigroup
    the unique idempotent in the subsemigroup generated by $x$.

    The set of locally threshold testable
   semigroups forms a pseudovariety of semigroups ($\cite {W}$, $\cite{BP}$).
So the theorem $\ref {0.0}$ implies
   \begin{cor} $\label {0.1}$
The pseudovariety of locally threshold testable semigroups consists
    of aperiodic semigroups and satisfies the pseudoidentity
    \[
    x^{\omega}yz^{\omega}ux^{\omega}tz^{\omega}=
    x^{\omega}tz^{\omega}ux^{\omega}yz^{\omega}
    \]
   \end{cor}

   \begin{lem} $\label {1.1}$
  Let the node ($\bf p, q$) be an $SCC$-node of $\Gamma^2$
of a locally threshold testable $DFA$ with
 state transition graph $\Gamma$ and suppose that ${\bf p} \sim \bf q$.

 Then ${\bf p} = \bf q$.
   \end{lem}
    Proof. The transition semigroup $S$ of the automaton is finite and
  aperiodic $\cite {L}$.  Suppose that for some element $e \in S$ and for
 some states $\bf q$ and $\bf p$ from $SCC$ $X$ we have ${\bf q}e={\bf q}$
 and ${\bf p}e=\bf p$.  In view of ${\bf q}e^i={\bf q}$, ${\bf p}e^i=\bf p$
and finiteness of $S$ we can assume $e$ is an idempotent.  In the $SCC$ $X$ for
some $a$, $b$ from $S$ we have ${\bf p}a={\bf q}$ and ${\bf q}b=\bf p$.
Hence, ${\bf p}eae={\bf q}$, ${\bf q}ebe=\bf p$.  So ${\bf p}eaebe={\bf
p}={\bf p}(eaebe)^i$ for any integer $i$.  There exists a natural number $n$
such that in the aperiodic semigroup $S$ we have $(eae)^n=(eae)^{n+1}$. From
 theorem $\ref {0.0}$ it follows that for the idempotent $e$, $eaeebe=ebeeae$.
 We have ${\bf p}={\bf p}eaebe={\bf p}(eaeebe)^n=
{\bf p}(eae)^n(ebe)^n={\bf p}(eae)^{n+1}(ebe)^n=
 {\bf p}(eae)^n(ebe)^neae= {\bf p}eae={\bf q}$.  So ${\bf p}={\bf q}$.
    $\Box$

 \begin{thm} $\label{4}$
 For $DFA$ {\bf A} with state transition graph $\Gamma$
   the following three conditions are equivalent:

  1){\bf A} is locally threshold testable.

  2)If the nodes (${\bf p,q}_1, {\bf r}_1$) and (${\bf q,r,t}_1, {\bf t}$) are
 $SCC$-nodes of $\Gamma^3$  and $\Gamma^4$, correspondingly, and

 $({\bf q,r})\succeq ({\bf q}_1,{\bf r}_1)$,
    $({\bf p,q}_1) \succeq ({\bf r,t})$,
    $({\bf p,r}_1) \succeq ({\bf q, t}_1)$ holds in $\Gamma^2$

  then ${\bf t}={\bf t}_1$.

  3)If the node (${\bf u,v}$) is an $SCC$-node of the graph $\Gamma^2$
 and ${\bf u} \sim {\bf v}$ then ${\bf u} = \bf v$.

 If the nodes (${\bf p,q}_1, {\bf r}_1$), (${\bf q,r,t}$),
 (${\bf q,r,t}_1)$  are $SCC$-nodes of the graph $\Gamma^3$ and

 $({\bf q,r})\succeq ({\bf q}_1,{\bf r}_1)$,
 $({\bf p,q}_1) \succeq ({\bf r, t})$,
 $({\bf p,r}_1) \succeq ({\bf q, t}_1)$ hold in $\Gamma^2$,

  then ${\bf t} \sim {\bf t}_1$.
   \end{thm}

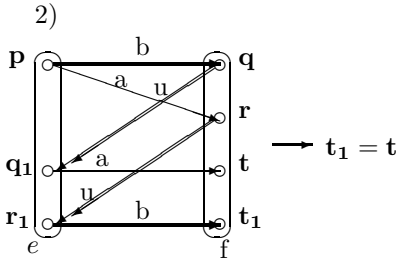
\begin{picture}(200,100)

\put(5,10){\circle{4}} \put(-3,-1){$e$}
 \put(7,10){\vector(1,0){63}}
 \put(7,9){\vector(1,0){63}}
    \put(-11,10){$\bf r_1$}

\put(70,10){\circle{4}} \put(77,10){$\bf t_1$}
 \put(70,-3){f}
  \put(38,12){b}

\put(5,30){\circle{4}}
 \put(7,30){\vector(1,0){63}}
\put(70,30){\circle{4}} \put(77,30){$\bf t$}
  \put(23,32){a}
  \put(-11,30){$\bf q_1$}
 \put(68,50){\vector(-3,-2){60}}  \put(77,50){$\bf r$}
 \put(68,49){\vector(-3,-2){54}}
\put(70,50){\circle{4}} \put(17,19){u}

\put(70,70){\circle{4}}  \put(77,70){$\bf q$}
 \put(7,70){\vector(1,0){63}}
 \put(7,71){\vector(1,0){63}}
\put(5,70){\circle{4}} \put(-10,71){$\bf p$}
  \put(38,74){b}
 \put(68,70){\vector(-3,-2){60}}
 \put(68,69){\vector(-3,-2){54}}
 \put(45,59){u}
 \put(7,70){\vector(3,-1){63}}
  \put(30,62){a}

\put(5,40){\oval(10,70)}
\put(69,40){\oval(10,70)}

 \put(90,40){\vector(1,0){15}}
  \put(110,36){$\bf t_1=t$}

  \put(0,86){$ 2) $}

 \end{picture}
\begin{picture}(200,100)

\put(5,10){\circle{4}} \put(-3,-1){$e$}
 \put(7,10){\vector(1,0){63}}
 \put(7,9){\vector(1,0){63}}
    \put(-11,10){$\bf r_1$}

\put(70,10){\circle{4}} \put(77,8){$\bf t_1$}
 \put(70,-3){$f_1$}
  \put(38,12){b}

\put(5,30){\circle{4}}
 \put(7,30){\vector(1,0){74}}
\put(81,30){\circle{4}} \put(82,33){$\bf t$}
 \put(83,20){$f_2$}

  \put(23,32){a}
  \put(-11,30){$\bf q_1$}
 \put(68,50){\vector(-3,-2){60}}  \put(77,50){$\bf r$}
 \put(68,49){\vector(-3,-2){54}}
\put(70,50){\circle{4}} \put(17,19){u}

\put(70,70){\circle{4}}  \put(77,68){$\bf q$}
 \put(7,70){\vector(1,0){63}}
 \put(7,71){\vector(1,0){63}}
\put(5,70){\circle{4}} \put(-10,71){$\bf p$}
  \put(38,74){b}
 \put(68,70){\vector(-3,-2){60}}
 \put(68,69){\vector(-3,-2){54}}
 \put(45,59){u}
 \put(7,70){\vector(3,-1){63}}
  \put(30,62){a}

\put(5,40){\oval(10,70)}
\put(69,40){\oval(10,70)}
\put(79,50){\oval(30,50)}

 \put(105,40){\vector(1,0){15}}
  \put(125,36){$\bf t_1 \sim t$}

  \put(0,86){$ 3) $}

 \end{picture}

  Proof.
 $2) \to 1)$:

  Let us consider the nodes
    ${\bf z}ebfueaf$ and ${\bf z}eafuebf$ where
   ${\bf z}$ is an arbitrary node of $\Gamma$, $a$, $u$, $b$ are arbitrary
  elements from transition semigroup $S$ of the automaton and $e$, $f$ are
 arbitrary idempotents from $S$. Let us denote

    ${\bf z}e={\bf p}$,
   ${\bf z}ebf={\bf q}$,
   ${\bf z}eaf={\bf r}$,
   ${\bf z}eafue={\bf r}_1$,
   ${\bf z}ebfue={\bf q}_1$,
   ${\bf z}ebfueaf={\bf t}$,
   ${\bf z}eafuebf={\bf t}_1$.

 By condition 2), we have ${\bf t}={\bf t}_1$, whence
   ${\bf z}ebfueaf={\bf z}eafuebf$.
  Thus, the condition $eafuebf=ebfueaf$ ($\ref {1}$)
   holds for the transition semigroup $S$.
    By theorem $\ref {0.0}$, the automaton is locally threshold testable.

 $1) \to 3)$:

  If the node (${\bf u,v}$) belongs to some $SCC$ of the graph $\Gamma^2$
 and ${\bf u} \sim {\bf v}$ then by lemma $\ref {1.1}$
 local threshold testability implies ${\bf u} = \bf v$.

  The condition $eafuebf=ebfueaf$ (($\ref {1}$), theorem $\ref {0.0}$)
   holds for the transition semigroup $S$ of the automaton.
  Let us consider nodes ${\bf p, q, r,t},{\bf q}_1, {\bf r}_1, {\bf t}_1$
 satisfying the condition 3). Suppose

  \centerline{(${\bf p,q}_1, {\bf r}_1)e=({\bf p,q}_1, {\bf r}_1)$,
 (${\bf q,r,t})f_2=({\bf q,r,t})$, (${\bf q,r,t}_1)f_1=({\bf q,r,t}_1$)}

 for some idempotents $e, f_1, f_2 \in S$, and

 \centerline{ $({\bf p,q}_1)a=({\bf r, t})$,
  $({\bf p,r}_1)b=({\bf q, t}_1)$
  $({\bf q,r})u=({\bf q}_1,{\bf r}_1)$}
 for some elements $a, b, u \in S$.
  Then ${\bf p}eaf_2={\bf p}eaf_1$ and ${\bf p}ebf_2={\bf p}ebf_1$.

  We have ${\bf t}_1f_2={\bf p}eaf_1uebf_1f_2$.
 By theorem $\ref {0.0}$,
   ${\bf p}ebf_jueaf_j={\bf p}eaf_juebf_j$ for $j=1,2$.
  So we have
${\bf t}_1f_2={\bf p}eaf_1uebf_1f_2={\bf p}ebf_1ueaf_1f_2$.
In view of ${\bf p}ebf_2={\bf p}ebf_1$ and $f_i=f_if_i$ we have
${\bf t}_1f_2={\bf p}ebf_2f_2ueaf_1f_2$. By theorem $\ref {0.0}$,
${\bf t}_1f_2={\bf p}e(bf_2)f_2ue(af_1)f_2={\bf p}e(af_1)f_2ue(bf_2)f_2$.
Now in view of ${\bf p}eaf_2={\bf p}eaf_1$ let us exclude $f_1$ and obtain
${\bf t}_1f_2={\bf p}eaf_2uebf_2={\bf t}$.
  So ${\bf t}_1f_2 ={\bf t}$. Analogously, ${\bf t}f_1={\bf t}_1$.

   Hence, ${\bf t}_1 \sim {\bf t}$.
  Thus 3) is a consequence of 1).

 $3) \to 2)$:

   Suppose that
   $({\bf p, q}_1,{\bf r}_1)e=({\bf p, q}_1,{\bf r}_1)$,
   (${\bf q, r, t, t}_1)f=({\bf q, r, t, t}_1)$,
   for some idempotents $e$, $f$
from transition semigroup $S$ of the automaton and

   \centerline{
   $({\bf p,q}_1)a=({\bf r, t})$,
     $({\bf p,r}_1)b=({\bf q, t}_1)$,
  $({\bf q,r})u=({\bf q}_1,{\bf r}_1)$}
  for some elements $a$, $u$, $b \in S$.
Therefore

 \centerline{
   $({\bf p,q}_1)eaf=({\bf p,q}_1)af=({\bf r, t})$}
   \centerline{
   $({\bf p,r}_1)ebf=({\bf p,r}_1)bf=({\bf q, t}_1)$}
   \centerline{
$({\bf q,r})u=({\bf q,r})fue=({\bf q}_1,{\bf r}_1)$}
   for idempotents $e$, $f$ and elements $a$, $u$, $b \in S$.

 For  $f=f_1=f_2$ from 3) we have
    ${\bf t} \sim {\bf t}_1$.
 Notice that $({\bf t}_1, {\bf t})f=({\bf t}_1, {\bf t})$.
  The node (${\bf t}_1, {\bf t}$) belongs to some $SCC$ of the graph
 $\Gamma^2$ and ${\bf t} \sim {\bf t}_1$, whence by by lemma $\ref {1.1}$,
${\bf t} = {\bf t}_1$.
   $\Box$

\begin{lem} $\label{1.7}$
  Let the nodes (${\bf q, r, t}_1$) and (${\bf q, r, t}_2$) be $SCC$-nodes
   of the graph $\Gamma^3$ of a locally threshold testable $DFA$ with state
transition graph $\Gamma$.
     Suppose that $({\bf p,r_1}) \succeq ({\bf q, t}_1)$,
   $({\bf p,r_1}) \succeq ({\bf q, t}_2)$ in the graph $\Gamma^2$
and ${\bf p} \succeq  {\bf r} \succeq {\bf r}_1$.

   Then ${\bf t}_1 \sim {\bf t}_2$.
   \end{lem}

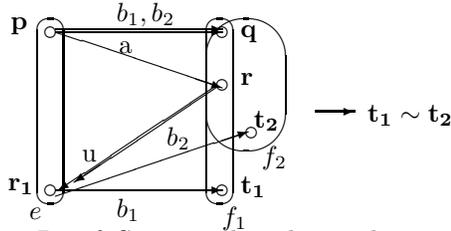
\begin{picture}(200,80)

\put(5,10){\circle{4}} \put(-3,-1){$e$}
 \put(7,10){\vector(1,0){63}}
    \put(-11,10){$\bf r_1$}

\put(70,10){\circle{4}} \put(77,9){$\bf t_1$}
 \put(70,-3){$f_1$}
  \put(30,0){$b_1$}
  \put(49,27){$b_2$}

\put(81,32){\circle{4}} \put(82,34){$\bf t_2$}
 \put(85,20){$f_2$}
 \put(7,8){\vector(3,1){73}}

 \put(68,50){\vector(-3,-2){60}}  \put(77,50){$\bf r$}
 \put(68,49){\vector(-3,-2){54}}
\put(70,50){\circle{4}} \put(17,21){u}

\put(70,70){\circle{4}}  \put(77,68){$\bf q$}
 \put(7,70){\vector(1,0){63}}
 \put(7,71){\vector(1,0){63}}
\put(5,70){\circle{4}} \put(-10,71){$\bf p$}
  \put(30,74){$b_1,b_2$}
 \put(7,70){\vector(3,-1){63}}
  \put(31,62){a}

\put(5,40){\oval(10,70)}
\put(69,40){\oval(10,70)}
\put(79,50){\oval(30,50)}

 \put(105,40){\vector(1,0){15}}
  \put(125,36){$\bf t_1 \sim t_2$}

 \end{picture}

  Proof. Suppose that the conditions of the lemma hold but
${\bf t}_1 \not\sim {\bf t}_2$.

   We have
   $({\bf p, r}_1)e=({\bf p,r}_1)$,
   (${\bf q,r,t}_1)f_1=({\bf q,r,t}_1)$,
   (${\bf q,r,t}_2)f_2=({\bf q,r,t}_2)$,
   for some idempotents $e$, $f_2$, $f_2$
from the transition semigroup $S$ of the automaton and

   \centerline{
   $({\bf p,r}_1)b_1=({\bf q, t}_1)$,
     $({\bf p,r}_1)b_2=({\bf q, t}_2)$,
    ${\bf p}a={\bf r}$,
    ${\bf r}u={\bf r}_1$}
  for some elements $a$, $u$, $b_1$, $b_2 \in S$.

If ${\bf t_1}f_2 \sim {\bf t}_2$
and ${\bf t_2}f_1 \sim {\bf t}_1$
then ${\bf t_2} \sim {\bf t}_1$ in spite of our assumption.
Therefore let us assume for instance that ${\bf t_1} \not\sim {\bf t}_2f_1$.
(And so ${\bf t_1} \neq {\bf t}_2f_1$).
This gives us an opportunity to consider ${\bf t}_2f_1$ instead of ${\bf t}_2$.
So let us denote ${\bf t}_2={\bf t}_2f_1$, $f=f_1=f_2$. Then
 ${\bf t}_2f={\bf t}_2$, ${\bf t}_1f={\bf t}_1$ and
${\bf t_1} \not\sim {\bf t}_2$.
Now

 \centerline{
   ${\bf p}eaf={\bf p}af={\bf r}$}
   \centerline{
   $({\bf p,r}_1)eb_1f=({\bf p,r}_1)b_1f=({\bf q, t}_1)$}
   \centerline{
   $({\bf p,r}_1)eb_2f=({\bf p,r}_1)b_2f=({\bf q, t}_2)$}
   \centerline{
    ${\bf r}u={\bf r}ue={\bf r}_1$}

So we have

\begin{picture}(200,85)

\put(5,10){\circle{4}} \put(-3,-1){$e$}
 \put(7,10){\vector(1,0){63}}
    \put(-11,10){$\bf r_1$}

\put(70,10){\circle{4}} \put(77,9){$\bf t_1$}
 \put(70,-3){$f$}
  \put(30,0){$b_1$}
  \put(49,27){$b_2$}

\put(81,32){\circle{4}} \put(82,36){$\bf t_2$}
 \put(85,20){$f$}
 \put(7,8){\vector(3,1){73}}

 \put(68,50){\vector(-3,-2){60}}  \put(77,50){$\bf r$}
 \put(68,49){\vector(-3,-2){54}}
\put(70,50){\circle{4}} \put(17,21){u}

\put(70,70){\circle{4}}  \put(77,68){$\bf q$}
 \put(7,70){\vector(1,0){63}}
 \put(7,71){\vector(1,0){63}}
\put(5,70){\circle{4}} \put(-10,71){$\bf p$}
  \put(30,74){$b_1,b_2$}
 \put(7,70){\vector(3,-1){63}}
  \put(31,62){a}

\put(5,40){\oval(10,70)}
\put(69,40){\oval(10,70)}
\put(79,50){\oval(30,50)}

 \put(115,40){$ and $}
  \put(145,40){$\bf t_1 \not\sim t_2$}

 \end{picture}

  Let us denote ${\bf q}_1={\bf q}ue$ and ${\bf t}={\bf q}_1af_1$.
Then

  \centerline{(${\bf p,q}_1, {\bf r}_1)e=({\bf p,q}_1, {\bf r}_1)$,
 (${\bf q,r})ue=({\bf q}_1,{\bf r}_1)$,
 (${\bf q,r,t,t}_i)f=({\bf q,r,t,t}_i)$}

  So the node (${\bf p,q}_1, {\bf r}_1$) is an $SCC$-node of the graph
$\Gamma^3$,
   the nodes (${\bf p,q, r,t}_i$) are $SCC$-nodes of the graph $\Gamma^4$
for $i=1,2$ and we have
 $({\bf q,r})\succeq ({\bf q}_1,{\bf r}_1)$,
 $({\bf p,q}_1) \succeq ({\bf r, t})$ and
 $({\bf p,r}_1) \succeq ({\bf q, t}_i)$ for $i=1,2$.

 Therefore, by theorem $\ref {4}$, (2), we have
${\bf t}_1={\bf t}$ and ${\bf t}_2={\bf t}$.
Hence, ${\bf t}_1 \sim {\bf t}_2$, contradiction.
   $\Box$
 \begin{defn} $\label{d}$
For any four nodes ${\bf p, q, r, r}_1$ of the graph $\Gamma$ of a
 $DFA$ such that
${\bf p} \succeq {\bf r} \succeq {\bf r}_1$, ${\bf p} \succeq {\bf q}$
and the nodes (${\bf p, r}_1$), (${\bf q, r}$) are $SCC$-nodes,
let $T_{SCC}({\bf p, q, r, r}_1)$ be the $SCC$ of $\Gamma$ containing
   \centerline{
$T({\bf p, q, r, r}_1):=\{ t$ $| ({\bf p,r}_1) \succeq ({\bf q, t})$
   and (${\bf q, r, t}$) is an $SCC$-node$\}$}
 \end{defn}

\begin{picture}(200,80)

\put(5,10){\circle{4}} \put(-3,-1){$e$}
 \put(7,10){\vector(1,0){63}}
    \put(-11,10){$\bf r_1$}

\put(70,10){\circle{4}} \put(77,9){$\bf t$}
 \put(70,-3){$f$}
  \put(30,0){$b$}

 \put(68,50){\vector(-3,-2){60}}  \put(77,50){$\bf r$}
 \put(68,49){\vector(-3,-2){54}}
\put(70,50){\circle{4}} \put(17,21){u}

\put(70,70){\circle{4}}  \put(77,68){$\bf q$}
 \put(7,70){\vector(1,0){63}}
 \put(7,71){\vector(1,0){63}}
\put(5,70){\circle{4}} \put(-10,71){$\bf p$}
  \put(30,74){$b$}
 \put(7,70){\vector(3,-1){63}}
  \put(31,62){a}

\put(5,40){\oval(10,70)}
\put(69,40){\oval(10,70)}

 \put(107,40){\vector(1,0){10}}
  \put(135,36){${\bf t} \in T_{SCC}({\bf p, q, r, r}_1)$}

 \end{picture}

In virtue of lemma $\ref {1.7}$, the $SCC$ $T_{SCC}({\bf p, q, r, r}_1)$
 of a locally threshold testable $DFA$ is well defined
(but empty if the set $T({\bf p, q, r, r}_1)$ is empty).
 Lemma $\ref {1.7}$ and theorem $\ref {4}$ (3) imply the following theorem

 \begin{thm} $\label{5}$
 A $DFA$ {\bf A} with state transition graph $\Gamma$
 is locally threshold testable iff

1)for every $SCC$-node ($\bf p, q$) of $\Gamma^2$ ${\bf p} \sim \bf q$
implies ${\bf p} = \bf q$

and

2)for every five nodes ${\bf p, q, r, q}_1, {\bf r}_1$ of the graph $\Gamma$
 such that
\begin{itemize}
 \item the non-empty $SCC$ $T_{SCC}({\bf p,q,r,r}_1)$ and
 $T_{SCC}({\bf p,r,q,q}_1)$ exist,
\item the node (${\bf p, q}_1,{\bf r}_1$) is an $SCC$-node of the
 graph $\Gamma^3$,
\item $({\bf q,r})\succeq ({\bf q}_1,{\bf r}_1)$ in $\Gamma^2$,
\end{itemize}
 holds $T_{SCC}({\bf p,q,r,r}_1)=T_{SCC}({\bf p,r,q,q}_1)$.
 \end {thm}

\begin{picture}(200,90)

\put(5,10){\circle{4}} \put(-3,-1){$e$}
 \put(7,10){\vector(1,0){63}}
 \put(7,9){\vector(1,0){63}}
    \put(-11,10){$\bf r_1$}

\put(70,10){\circle{4}}
  \put(77,8){${\bf t}_1 \in T_{SCC}({\bf p, r, q, q}_1)$}
 \put(70,-3){$f_1$}
  \put(38,12){b}

\put(5,30){\circle{4}}
 \put(7,30){\vector(1,0){74}}
\put(81,30){\circle{4}}
  \put(85,31){${\bf t} \in T_{SCC}({\bf p, q, r, r}_1)$}
 \put(83,20){$f_2$}

  \put(23,32){a}
  \put(-11,30){$\bf q_1$}
 \put(68,50){\vector(-3,-2){60}}  \put(77,50){$\bf r$}
 \put(68,49){\vector(-3,-2){54}}
\put(70,50){\circle{4}} \put(17,19){u}

\put(70,70){\circle{4}}  \put(77,68){$\bf q$}
 \put(7,70){\vector(1,0){63}}
 \put(7,71){\vector(1,0){63}}
\put(5,70){\circle{4}} \put(-10,71){$\bf p$}
  \put(38,74){b}
 \put(68,70){\vector(-3,-2){60}}
 \put(68,69){\vector(-3,-2){54}}
 \put(45,59){u}
 \put(7,70){\vector(3,-1){63}}
  \put(30,62){a}

\put(5,40){\oval(10,70)}
\put(69,40){\oval(10,70)}
\put(79,50){\oval(30,50)}

 \put(195,40){\vector(1,0){15}}
  \put(225,40){$T_{SCC}({\bf p, q, r, r}_1)=T_{SCC}({\bf p, r, q, q}_1)$}

 \end{picture}

     \section{Algorithm to verify the local threshold testability}
   A linear depth-first search algorithm finding
 all $SCC$ of the given directed graph
(see $\cite {A}$, $\cite {Ta}$ or $\cite {K94}$) will be used.

  \subsection{To check the reachability on an oriented graph}

 For a given node ${\bf q_0}$, we consider depth-first
 search from the node. First only ${\bf q_0}$ will be marked.
 Every edge is crossed two times.
 Given a node, the considered path includes
 first the ingoing edges and then the outgoing edges.
  After crossing an edge  in the positive direction
 from the marked node ${\bf q}$ to the node $\bf r $ we mark $\bf r$ too.
 The process is linear in the number of edges
(see $\cite {A}$, $\cite {K89}$ for details).

 The set of marked nodes forms a set of nodes that are reachable from
 ${\bf q_0}$. The procedure may be repeated for any node of the graph $G$.

 The time of the algorithm for all pairs of nodes is $O(n^2)$.

  \subsection{To verify local threshold testability}

    Let us find all $SCC$ of the graphs $\Gamma$, $\Gamma^2$ and $\Gamma^3$
 and mark all $SCC$-nodes ($O(n^3)$ time complexity).

   Let us recognize the reachability on the graph $\Gamma$ and
  $\Gamma^2$ and form the table of reachability for all pairs of $\Gamma$ and
  $\Gamma^2$.  The time required for this step is $O(n^4)$.

   Let us check the conditions of lemma {\ref {1.1}}.
  For every $SCC$-node ($\bf p, q$) (${\bf p} \neq \bf q$) from $\Gamma^2$
 let us check
  the condition ${\bf p} \sim \bf q$.  A negative answer for any considered
 node ($\bf p, q$) implies the validity of the condition. In opposite case
 the automaton is not locally threshold testable.
   The time of the step is $O(n^2)$.

  For every four nodes ${\bf p, q, r, r}_1$ of the graph
$\Gamma$, let us check the following conditions (see {\ref {d}}):
${\bf p} \succeq {\bf r} \succeq {\bf r}_1$ and ${\bf p} \succeq {\bf q}$.
In a positive case, let us form $SCC$ $T_{SCC}({\bf p, q, r, r}_1)$ of all
 nodes ${\bf t} \in \Gamma$ such that
 $({\bf p,r}_1) \succeq ({\bf q, t})$ and (${\bf q, r, t}$) is an $SCC$-node.
 In case that $SCC$ $T_{SCC}$ is not well defined the automaton is not
 threshold testable.
   The time required for this step is $O(n^5)$.

 For every five nodes ${\bf p,q,r,q}_1, {\bf r}_1$ from
$\Gamma$ let us check now the second condition of theorem {\ref {5}}.
If non-empty components $T_{SCC}({\bf p,q,r,r}_1)$ and
 $T_{SCC}({\bf p,r,q,q}_1)$ exist,
the node (${\bf p, q}_1,{\bf r}_1$) is an $SCC$-node of the graph $\Gamma^3$
and $({\bf q,r})\succeq ({\bf q}_1,{\bf r}_1)$ in $\Gamma^2$,
let us verify the equality $T_{SCC}({\bf p,q,r,r}_1)=T_{SCC}({\bf p,r,q,q}_1)$.
 If the answer is negative then the automaton is not threshold testable.
  A positive answer for all considered cases implies the validity of the
 condition of the theorem. The time required for this step is $O(n^5)$.

The whole time of the algorithm to check the local threshold
testability is $O(n^5)$.

\section{The local testability}
  We present now necessary and sufficient conditions of local testability of
 Kim, McNaughton and McCloskey ($\cite {K89}$, $\cite {K91}$)
 in the following form:
 \begin{thm} $\label {5.1}$ ($\cite {K91}$)
  A $DFA$ with state transition graph $\Gamma$ and transition semigroup $S$
 is locally testable iff the following two conditions hold:

  1)For any  $SCC$-node ($\bf p, q$) from $\Gamma^2$
   such that ${\bf p} \sim \bf q$ we have ${\bf p} = \bf q$.

 2)For any  $SCC$-node ($\bf p, q$) from $\Gamma^2$ such that
 ${\bf p} \succ \bf q$ and arbitrary element $s$ from $S$
  we have
${\bf p}s \succeq \bf q$ is valid iff ${\bf q}s \succeq \bf q$.
   \end{thm}
  The theorem implies

 \begin{cor} $\label {5.2}$
  A $DFA$ with state transition graph $\Gamma$ over alphabet $\Sigma$
 is locally testable iff the following two conditions hold:

  1)For any  $SCC$-node ($\bf p, q$) from $\Gamma^2$
   such that ${\bf p} \sim \bf q$ we have ${\bf p} = \bf q$.

 2)For any node ($\bf r, s$) and any  $SCC$-node ($\bf p, q$) from
$\Gamma^2$ such that  $({\bf p, q}) \succ (\bf r, s)$,
 ${\bf s} \sim \bf q$ and for arbitrary $\sigma$ from $\Sigma$ we have
${\bf r}\sigma \succeq \bf s$ is valid iff ${\bf s}\sigma \succeq \bf s$.
   \end{cor}

     \section{Algorithm to verify the local testability}

   In  $\cite {K91}$, a polynomial time algorithm for local testability
problem was considered.
 Now we present another version of such algorithm with the same time
complexity. We hope that it will be more simple.

 Let us form a table of reachability on the graph $\Gamma$
 ($O(n^2)$ time complexity).

 Let us find $\Gamma^2$ and all $SCC$-nodes of $\Gamma^2$.

  For every $SCC$-node ($\bf p, q$) (${\bf p} \neq \bf q$) from $\Gamma^2$
let us check the condition ${\bf p} \sim \bf q$. ($O(n^2)$ time complexity).
If the condition holds then the automaton is not locally testable
($\ref {5.2}$).

     Let us exclude all edges ($\bf p, q) \to (\bf r, s$) from the graph
 $\Gamma^2$ such that ${\bf s} \not\succeq \bf q$ and
${\bf s} \not\succeq \bf p$.
Then let us mark all nodes ($\bf p, q$) of the graph $\Gamma^2$
such that for some $\sigma$ from $\Sigma$ from the two conditions
${\bf p}\sigma \succeq \bf q$ and ${\bf q}\sigma \succeq \bf q$
only one is valid.
The time required for this step is $O(n^2)$.

  Then we add to the graph new node ($\bf 0, 0$) with edges from this node
 to every $SCC$-node.
 Let us find the set of nodes reachable from the node ($\bf 0, 0$).
 ($O(n^2)$ time complexity).
  The automaton is locally testable iff no marked node belongs to obtained
set  ($\ref {5.2}$).

 The whole time of the algorithm to check the local testability is $O(n^2)$.

\section*{Acknowledgments}
  I would like to express my gratitude to Stuart Margolis for posing the
  problem and for helpful suggestions on improving the style of the paper
 and to referees for important and useful remarks.

 \end{document}